\documentclass[reprint,superscriptaddress,showpacs,preprintnumbers,amsmath,amssymb,aps,prl]{revtex4-1}

\usepackage{graphicx}
\usepackage{dcolumn}
\usepackage{bm}
\usepackage{hyperref}

\begin{document}

\title{Spatial distribution of phase singularities in optical random vector waves}

\author{L. De Angelis}
\affiliation{%
 Center for Nanophotonics, AMOLF, Science Park 104 1098 XG Amsterdam, The Netherlands
}%
\author{F. Alpeggiani}
\affiliation{%
 Center for Nanophotonics, AMOLF, Science Park 104 1098 XG Amsterdam, The Netherlands
}%
\author{A. Di Falco}
\affiliation{
SUPA, School of Physics and Astronomy, University of St Andrews, North Haugh, St Andrews KY16 9SS, UK
}
\author{L. Kuipers}
\email{kuipers@amolf.nl}
\affiliation{%
 Center for Nanophotonics, AMOLF, Science Park 104 1098 XG Amsterdam, The Netherlands
}%

\date{\today}

\begin{abstract}
Phase singularities are dislocations widely studied
in optical fields as well as in other areas of physics.
With experiment and theory we show
that the vectorial nature of light affects
the spatial distribution of phase singularities in random light fields.
While in scalar random waves phase singularities exhibit spatial
distributions reminiscent of particles in isotropic liquids, in vector fields 
their distribution for the different
vector components becomes anisotropic
due to the direct relation between propagation
and field direction.
By incorporating this relation in the theory
for scalar fields by Berry and Dennis,
we quantitatively describe our experiments.
\end{abstract}


\maketitle

%%%%%%%%%%%%%%%%%%%%%%%%%%%%%%%%%%%%%%%%%%%%%%%%%%%%%%%%%%%%%%%%%%%%%
%%\section{Introduction}

Finding correlations in chaotic systems is the first step towards understanding.
Many are the fields where such predictions could be exploited, from weather forecast to
economic modeling \cite{wilks99,laloux99}.
The study of random phenomena is a topic of great interest and
inspiration for many branches of physics as well.
In electromagnetism, for example, random wave fields have been a topic of intense
studies since decades, an outstanding example being Anderson localization 
of light~\cite{Lagendijk2009}. 
More recently the scientific interest
on random wave fields has continued intensively, ranging from
useful techniques as non-invasive imaging with
speckle correlation~\cite{Bertolotti2012} to fascination
concerning the observation of rogue waves
in optical fields~\cite{rogue,Delre15}.
Zooming into the structure of a random wave field,
attention has been pointed to
deep-subwavelength dislocations known as phase singularities~\cite{nyeberry}.

Phase singularities are locations
in which the phase of a scalar complex field is not defined.
In two-dimensional fields these locations are points in the plane.
Although they are just a discrete set of points,
phase singularities can describe the basic properties
of the field in which they arise.
For this reason they are widely studied in wave fields 
\cite{freund93,freund95,Curtis03,palacios04,Vignolini10,Peano15,rotenberg2015},
as well as in many areas of physics,
where they are better-known as topological
defects in nematics~\cite{fernandez2007}
or as vortices in superfluids~\cite{superf}.

For a single frequency phase singularities are fixed in space, and
their spatial distribution
in a scalar field of monochromatic random waves
has been analytically modeled by Berry and Dennis 
\cite{berrydennis}.
The hallmark of such a distribution is a
clear pair correlation,
reminiscent of that of particles in liquids. 
By realizing random waves ensembles
in microwave billiards 
\cite{stockmann2002,kim2005measurement,stockmann2006quantum},
the correlation of phase 
singularities was tested for a field perpendicular
to the plane of the billiard, 
showing excellent agreement with the 
theoretical expectations \cite{stockmann2009}.
For such a field and
in that geometry
indeed scalar theory was appropriate.
However, electromagnetic waves are vectorial
in nature, and in a different framework it was already
demonstrated how the presence of a spin degree of freedom
can affect the correlation properties of a random 
field~\cite{urbina13,ngo01}.

%%%%%%%%%%%%%%%%%%%%%%%%%%%%%%%%%%%%%%%%%%%%%%%%%%%%%%%%%%%%%%%%%%%%%
%%\section{Here we}

Here, we show how the vectorial nature of light
affects the distribution of its phase singularities. 
By mapping the in-plane optical vector field measured
above a chaotic resonator
we investigate the distribution of phase singularities
in two-dimensional random vector waves.
We show that the distribution of phase singularities
deviates from that for scalar random waves.
This deviation is caused by the relation between
the transverse field
and waves propagation direction.
Thus, even when the considered vector field is equipartitioned with respect to
both the in-plane polarization and propagation direction, any specific choice of
field component directly leads to an anisotropic distribution of
the contributing propagation directions.
By treating this anisotropy with an analytical model we
quantitatively explain our experimental observations.
Finally, we show how an out-of-plane component
that we construct from our in-plane fields
obeys the predictions for scalar fields.

%%%%%%%%%%%%%%%%%%%%%%%%%%%%%%%%%%%%%%%%%%%%%%%%%%%%%%%%%%%%%%%%%%%%%
%%\section{Near-field measurements} 

\begin{figure}[t]
\centering
\includegraphics[width = 8.5cm]{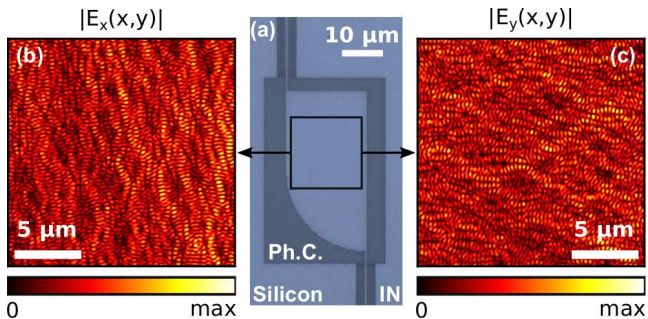}
\caption{Near-field measurement of the amplitude 
of the two cartesian component $E_x$ (b)
and $E_y$ (c) of the optical electric field in the
\emph{Chaotic Cavity} (optical micrograph shown) (a).}
\label{fig:meas}
\end{figure}

To generate optical random waves we inject monochromatic light
($\lambda_0 = 1550$ nm) in a chaotic 
cavity: a silicon-on-insulator membrane
enclosed by a planar photonic crystal
with one input and one output waveguide [Fig.~\ref{fig:meas}(a)].
Light is coupled through the input waveguide,
exciting a Transverse Electric (TE) slab mode in the silicon membrane,
which has its electric field vector in the membrane plane.

With our near-field microscope we map the in-plane components
of the optical field in the cavity, which is in fact the only non-vanishing
component of the electric field.
The near field is locally converted and
delivered to the far field through an optical fiber. A heterodyne
detection scheme enables the measurement of amplitude and phase
\cite{balistreri2000}.
With standard polarization optics we
selectively detect the $(E_x,E_y)$
Cartesian 
components of the in-plane electric field $\mathbf{E}$~\cite{Burresi09},
thus gaining access to its vectorial content.
By scanning the surface of the sample we measure a two-dimensional
map of the in-plane complex optical field above the cavity.

Figure \ref{fig:meas} presents the measurements of the amplitude of
$E_x$ and $E_y$.
The field pattern that results in the cavity
can be thought of as a random superposition
of plane waves \cite{stockmann2006quantum}. We find that for the
$x$- and $y$-components of the field the probability density function of the intensity
is exponential \cite{supplemental}, underlying the randomness.
The near-field maps [Fig.~\ref{fig:meas}(b),(c)] depict 
the patterns resulting from
interference of light in the chaotic cavity.
At first sight destructive and constructive interference occurs
at random locations in the plane. On closer inspection a 
difference between the maps of the two field components catches the eye:
the features of each pattern exhibit a preferred axis, 
related to the specific field component.
A vertical stripy pattern 
with roughly $1-2$ micrometers between the stripes
is present in the $E_x$ field. A modulation
of the amplitude is observed along the stripes as well,
characterized this time by a shorter
length scale.
The case of $E_y$ is completely analogous, but rotated by
90 degrees.

%%%%%%%%%%%%%%%%%%%%%%%%%%%%%%%%%%%%%%%%%%%%%%%%%%%%%%%%%%%%%%%%%%%%%
%%\section{phase singularities pinpointing}

\begin{figure}[t]
\centering
\includegraphics[width = 8.5cm]{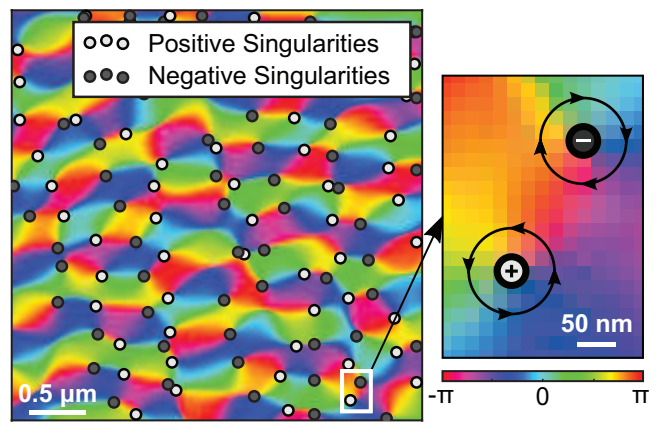}
\caption{False-color map of the measured phase of $E_x$. Phase singularities are
represented (circles) with their topological charge: +1 (positive) or
-1 (negative). The zoom highlights how the direction of the 
circulation of the phase around the 
singular point determines its topological charge.}
\label{fig:phasesing}
\end{figure}

By separately measuring the Cartesian components
of the electric field we implicitly
established a criterion to depict a vector
${\bf E}$ by using two scalar quantities $(E_x,E_y)$,
in which we can now seek for phase singularities \cite{berryintro}.
Please note that such singularities cannot be found
in the total intensity, which has no vanishing points.
In a two-dimensional scalar complex field $\psi({\bf r})$,
phase singularities are points in which
its phase $\varphi$ is undefined.
The phase circulates around the singular points,
assuming all its possible values from $-\pi$ to $\pi$
\cite{nyeberry}.
Quantitatively, the line integral of $\varphi$
along a path $\mathcal{C}$ enclosing only one 
singularity yields an integer multiple of $2\pi$:
\begin{equation}\label{eq:phasecirc}
\int_\mathcal{C}d\varphi = 2\pi s,
\end{equation}
where the integer $s$ is called \emph{topological charge} of the singularity.
The definition of topological charge also gives us
a powerful way to identify phase singularities. 
We calculate the integral of Eq.~\eqref{eq:phasecirc} along 
$2\times 2$ pixels loops at every point of the measured phase map, determining
position and topological charge of all the optical vortices in the field,
with a spatial accuracy that is limited by the pixel size,
of approximately $20$ nm.
Figure \ref{fig:phasesing} shows the phase singularities pinpointed
in a subset of the phase map of $E_x$.
Only topological charges of $\pm 1$ are observed.

%%%%%%%%%%%%%%%%%%%%%%%%%%%%%%%%%%%%%%%%%%%%%%%%%%%%%%%%%%%%%%%%%%%%%
%%\section{Distribution characterization}

The distribution of optical vortices in the plane is rather disordered
(Fig.~\ref{fig:phasesing}), although already by eye a spatial
correlation seems discernible if taking
into account the topological charge.
In order to to unveil such correlation
we quantitatively characterize their spatial distribution.
A natural way to do this is by calculating the
\emph{pair correlation function}
\begin{equation}
g(r) = \frac{1}{N\rho} \langle \sum_{i \neq j} \delta ( r -|{\bf r}_j - {\bf r}_i|) \rangle ,\label{eq:g_calc}
\end{equation}
where $N$ is the total number of singularities, $\rho$ is the average density 
of surrounding singularities and $\delta$ the Dirac function.
This function is directly related to the probability of finding
two entities (${\bf r}_i$ and ${\bf r}_j$) at a distance
$r$ from each other and is widely used in physics to describe 
discrete systems of various kinds \cite{liquids}.

\begin{figure}[t]
\centering
\includegraphics[width =8.6cm]{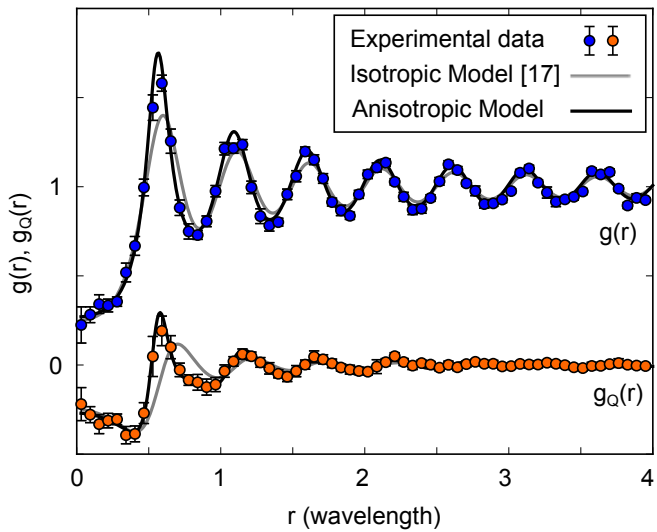}
\caption{Pair (blue) and Charge (orange) correlation function of
phase singularities in the measured field $E_x$. The gray
line is the theoretical expectation for a scalar field of isotropic
random waves \cite{berrydennis}. Our data significantly deviates from
such theory, while a perfect agreement is obtained by considering
a new model that includes directional anisotropies (black lines).}
\label{fig:gTEx}
\end{figure}

Figure \ref{fig:gTEx} presents the $g(r)$ calculated from
our experimental data,
specifically for the full data set of
singularities of $E_x$
\footnote{Similar results are observed for the case of the $E_y$ field.}.
The shape of the distribution function is highly similar 
to what is typically observed for a system of particles
in a liquid~\cite{liquids}. After an initial dip 
$g(r)$ oscillates around one, with an 
amplitude that decreases as $r$ is increased. 
The first peak, representative of a surplus of singularities, emerges
at a distance of roughly half a wavelength.
The decrease in amplitude of the oscillations describes
the loss of correlation of the system.
However, one peculiarity that we observe is definitely different
compared to the case of a liquid:
$g(r)$ approaches a finite value for $r\approx 0$.
This means that asymptotically there is a finite probability of finding
two vortices at the same location. While unusual this is in fact allowed
by the zero-dimensionality of optical vortices.

In analogy to what typically done for ionic liquids~\cite{liquids},
it is convenient to introduce a generalized expression of the
$g(r)$ for a system of \emph{charged} entities. In the \emph{charge
correlation function} $g_Q(r)$ each vortex is weighted with its topological
charge $s$:
\begin{equation}
g_Q(r) = \frac{1}{N\rho} \langle \sum_{i \neq j} \delta ( r -|\mathbf{r}_j- \mathbf{r}_i|) \,s_i\,s_j\rangle .\label{eq:gQ_calc}
\end{equation}
The experimental result for the $g_Q(r)$ is also reported in Fig.~\ref{fig:gTEx},
providing new information about our system.
The main observation here is
that $g_Q(r)$ is approximately equal to $-g(r)$ in the region $r\approx 0$,
meaning that only singularities with opposite topological
charge are likely to be indefinitely close to each other.
This behavior is usually
interpreted in terms of reciprocal screening among critical points with opposite
topological charge \cite{freund98,freund15}. Notably, critical-point screening is related to the reduction
of topological charge fluctuations inside a finite region with respect to the prediction for a collection of random charges \cite{freund98,berrydennis,vantiggelen06}.

In  an influential paper \cite{berrydennis} Berry and Dennis calculated
the correlation functions of singularities in a scalar field~$\psi$,
modeled as a superposition of plane waves with the same momentum and random phases $\delta_{\mathbf{k}}$, i.e.,
\begin{equation}
\psi(\mathbf{r}) = \sum_{\mathbf{k}} a_{k}
\exp(i \mathbf{k}\cdot\mathbf{r} + i \delta_{\mathbf{k}}).\label{eq:rwsuperpos}
\end{equation}
The model assumes that the waves amplitudes are isotropically distributed
along a circle of radius $k_0$ in Fourier space.
The results of this isotropic model are shown as solid gray lines in
Fig.~\ref{fig:gTEx}.
Most of the key features of the experimental distribution
are qualitatively accounted for by the model,
but we clearly observe some deviation
from the theory.
The biggest difference is in the $g_Q(r)$, where the first
peak turns out to be narrower than in theory,
as well as significantly shifted towards lower distances. 
This is in contrast with what was observed for out-of-plane fields
in microwaves billiards~\cite{stockmann2009}, where excellent agreement
was found.

%%%%%%%%%%%%%%%%%%%%%%%%%%%%%%%%%%%%%%%%%%%%%%%%%%%%%%%%%%%%%%%%%%%%%
%%\section{directional anisotropy in the field}

The origin of the observed discrepancies with respect to $g(r)$ and
$g_Q(r)$ lies in the vector nature of the light.
For the TE modes a direct relation exists between
the selected in-plane field component
and the direction of propagation:
the modes will have no electric field component along the direction of 
propagation. Therefore the choice of field component to be
investigated (e.g. $E_x$ in Fig.~\ref{fig:phasesing}) affects the distribution
of propagation directions that contribute to the wave pattern.
Whereas the general model in Eq.~\eqref{eq:rwsuperpos} remains valid, the
anisotropy of our system violates the assumption of isotropy \cite{berrydennis},
i.e., that $a_{k}$ only depends on the magnitude of $\mathbf{k}$. As a consequence,
the field correlation function must display an additional
dependence on the relative spatial orientation of the points.

We now calculate the correlation properties of the in-plane components of the 
field and of the corresponding singularities distribution, by including such anisotropy
in a modified version of the original model.
Since for a TE mode ${\bf E}(\mathbf{k}) \perp \mathbf{k}$,
the Fourier coefficients of the in-plane field components are effectively
modulated by the sine of the angle $\theta_{\mathbf{k}}$ enclosed by the
direction of the considered field component and the in-plane wavevector $\mathbf{k}$.
For this reason, we assume
\begin{equation}
E_{j}(\mathbf{r}) \propto \sum_{\mathbf{k}} \sin(\theta_{\mathbf{k}})
\exp(i \mathbf{k}\cdot\mathbf{r} + i \delta_{\mathbf{k}}),\quad j = x,y.
\label{eq:anis}
\end{equation}
Note that the total intensity, $E_x^2+E_y^2$ remains isotropic.
%%%%%%%%%%%%%%%%%%%%%%%%%%%%%%%%%%%%%%%%%%%%%%%%%%%%%%%%%%%%%%%%%%%%%
%%\section{Theory}

The additional angular dependence in the Fourier expansion of Eq.~\ref{eq:anis}
influences the correlation properties of the wave field.
In particular, the spatial autocorrelation function of each field component,
\begin{equation}
C({\bf r})
= \int \!\! d{\bf r}'\, E^*_j({\bf r}')E_j({\bf r}' + {\bf r})
= \frac{1}{2\pi} \int \!\! d{\bf k}\, 
\left| E_j({\bf k}) \right|^2\,  e^{-i{\bf k}\cdot{\bf r}},
\end{equation}
exhibits a dependence on the orientation $\varphi$ of vector $\mathbf{r}$:
\begin{eqnarray}
C({\bf r}) \doteq C(r, \varphi) & = & \frac{1}{2 \pi} \int \!\! 
d\theta_{\mathbf{k}}\,
\sin^2(\theta_\mathbf{k}) e^{-ik_0r\cos(\theta_\mathbf{k} - \varphi)} \\
              & = & \frac{1}{2}\left [ 
J_0(k_0 r) + \cos(2\varphi) J_2(k_0 r) \right ],
\label{eq:Cr_anis}\end{eqnarray}
where $J_n(x)$ is the Bessel function of order $n$ and $k_0$ is the wavenumber of the TE mode.
This is in contrast with the case of a fully isotropic scalar field,
where $C(\mathbf{r}) = J_0(k_0 r)$~\cite{berrydennis}.
The autocorrelation function of the field contains all the
information needed to retrieve pair and charge correlation
functions of the phase singularities \cite{supplemental}.
Analogously to $C(r,\varphi)$, the pair and charge correlation
functions, $g^{\mathrm{(an)}}(r,\varphi)$ and $g_Q^{\mathrm{(an)}}(r,\varphi)$,
display a dependence on the spatial vector orientation $\varphi$.
Since in the corresponding experimental quantities in
Eqs.~\eqref{eq:g_calc} and \eqref{eq:gQ_calc} the average is taken
over all reciprocal orientations of the points,
to compare the experimental data with the theoretical results, we average the latter over the polar angle:
\begin{equation}\label{eq:gQ_av}
g_Q^{\mathrm{(an)}}(r) = \frac{1}{2\pi}\int_0^{2\pi} d\varphi\; g_Q^{\mathrm{(an)}}(r,\varphi)
\end{equation}
[and similarly for $g^{\mathrm{(an)}}(\mathbf{r})$].
The black solid lines in Fig.~\ref{fig:gTEx} show the analytic
outcome of such calculations (Anisotropic Model).
Comparison between experiment and the new model now exhibits an excellent
qualitative and quantitative agreement.
This confirms that the anisotropy in the direction distribution of random waves,
intrinsic in the vector nature of optical wave fields,
significantly affects the spatial distribution of phase singularities.

%%%%%%%%%%%%%%%%%%%%%%%%%%%%%%%%%%%%%%%%%%%%%%%%%%%%%%%%%%%%%%%%%%%%%
%%\section{Directional anisotropy in gQ}

\begin{figure}[t]
\centering
\includegraphics[width = 8cm]{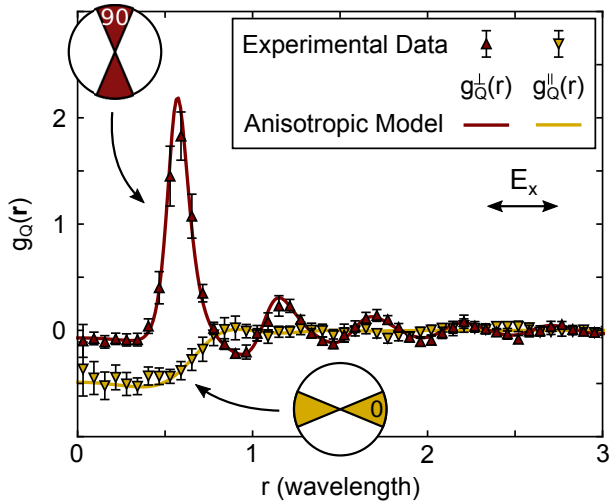}
\caption{Directional charge correlation function. We repoort 
the two illustrative cases of direction perpendicular
(red) or parallel (yellow) to the direction of the considered
field component ($E_x$). The distribution function strongly depends
on the direction in the plane. Experimental data is represented by
the triangles, while lines show our modified model
for anisotropic random waves, which perfectly fits the data.}
\label{fig:gQdir}
\end{figure}

A further confirmation of the validity of our model comes
from restricting the orientation of the spatial displacement
vector $\mathbf{r}$ to a limited range of values ($\Delta\varphi = \pi/4$)
around the directions perpendicular ($\perp$) and parallel
($\parallel$) to the field projection axis. In Fig.~\ref{fig:gQdir},
we compare the results for both the experimentally calculated
$g_Q(\mathbf{r})$ and the restricted averages of Eq.~\eqref{eq:gQ_av} (Anisotropic Model).
The two direction-dependent distribution functions show a
completely different behavior,
neither being equal to the isotropic $g_Q(r)$ of Ref.~\cite{berrydennis}, which, of course, does not display any orientation dependence.
Several differences can be spotted between $g^\parallel_Q$ and $g^\perp_Q$.
First, $g^\perp_Q(r)$ vanishes as $r$ approaches zero, while
$g^\parallel_Q(r)$ does not.
As a consequence singularities of opposite sign are most likely to be 
arbitrarily close along the polarization direction.
Secondly, in the $g^\perp_Q(r)$ there is an evident and positive peak,
followed a number of clear peaks with decreasing height. 
Sequences of same-sign
singularities spaced by approximately half a wavelength
are therefore expected along the 
direction perpendicular to the polarization (see also Fig.~\ref{fig:phasesing}).
Here the loss of spatial correlation is  slow
compared to the direction parallel 
to the polarization, along which any correlation
structure is immediately lost
after the initial dip in the $g^\parallel_Q(r \approx 0)$.

%%%%%%%%%%%%%%%%%%%%%%%%%%%%%%%%%%%%%%%%%%%%%%%%%%%%%%%%%%%%%%%%%%%%%
%%\section{Scalar field}

It is clear that the 
vector nature of the optical
electric field impacts the spatial distribution of phase singularities.
Interestingly, an out-of-plane field component would give us
access to a quantity that behaves like a scalar.
By Fourier transforming the measured complex fields $E_x$ and $E_y$
we can calculate the wave-vector space distribution of the magnetic field
$\mathbf{H} \propto  \mathbf{k} \times \mathbf{E}$.
Fourier transforming back, we can thus construct a spatial map of $H_z$
(up to a constant) \cite{Olmon10},
in which we identify
singularities and perform the same statistical
analysis done for $E_x$ and $E_y$.

\begin{figure}[b]
\centering
\includegraphics[width = 8.6cm]{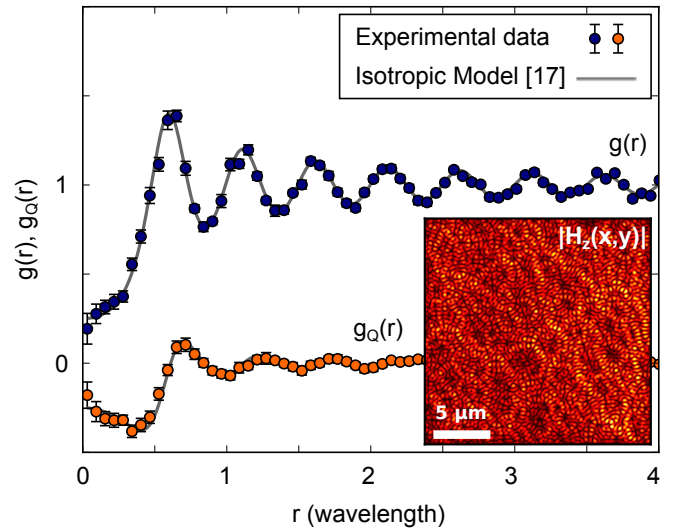}
\caption{Pair and charge correlation function of phase singularities
in the constructed out-of-plane magnetic field $H_z(x,y)$. The distribution functions for singularities
in this scalar field are
in perfect agreement with the model for isotropic random waves
\cite{berrydennis}. In the inset an amplitude map of $H_z(x,y)$ (a.u.).}
\label{fig:gTE}
\end{figure}

The inset in Fig.~\ref{fig:gTE} shows the amplitude of $H_z({\bf r})$
as constructed from our measured data.
No anisotropy is evident in the resulting amplitude map,
in contrast to what we observed for the constituent fields $E_x$ and $E_y$.
Figure \ref{fig:gTE} presents the distribution functions $g(r)$ and $g_Q(r)$ for the 
phase singularities in $H_z$, together with the theoretical
model for isotropic random waves \cite{berrydennis}.
The agreement is in this case excellent. The
direction-dependent distribution functions (not shown)
do not exhibit any anisotropy.

%%%%%%%%%%%%%%%%%%%%%%%%%%%%%%%%%%%%%%%%%%%%%%%%%%%%%%%%%%%%%%%%%%%%%
%%\section{Conclusion}

When considering phase singularities in optical fields one 
needs to take into account that light is in general described as a vector wave.
We studied the case in which the Cartesian components of an optical random
field are separately considered as scalar quantities.
We noticed how considering each field component goes hand in hand with 
directional anisotropies in the distribution of propagation direction of random waves.
This leads to significant consequences for the spatial distribution of optical vortices.
As discussed by analyzing experimental results
supported by analytical model, the differences become particularly
dramatic when considering the angular dependence of the distribution.
We stress that the anisotropic behavior that we analyzed in this Letter is a consequence
of the vector nature of light and it is not related to the shape or dielectric constituents
of the optical cavity that we used. For this reason, we believe that similar phenomena
should arise every time that a truly vectorial electromagnetic
field is projected along one of its component.

%%%%%%%%%%%%%%%%%%%%%%%%%%%%%%%%%%%%%%%%%%%%%%%%%%%%%%%%%%%%%%%%%%%%%
%%\section{Acknowledgments}

We thank Boris le Feber for useful discussions,
and Pieter Rein ten Wolde and A. Femius Koenderink
for critical reading of the manuscript.
This work is part of the research program of the Netherlands Foundation for Fundamental Research
on Matter (FOM) and the Netherlands Organization for Scientific Research (NWO). We acknowledge funding from ERC Advanced,
Investigator Grant (no. 240438-CONSTANS).
ADF acknowledges support from EPSRC (EP/L017008/1).

%%%%%%%%%%%%%%%%%%%%%%%%%%%%%%%%%%%%%%%%%%%%%%%%%%%%%%%%%%%%%%%%%%%%%
%%\section{Bibliography}


\end{document}